\documentclass[12pt]{article}
\usepackage[centertags]{amsmath}
\usepackage{amsfonts}
\usepackage{graphicx}
\usepackage[]{epsfig}
\usepackage{amssymb}
\newcommand{\id}{i\kern.06em\hbox{\raise.25ex\hbox{$/$}\kern-.60em$\partial$}}
\newcommand{\dd}{\kern.06em\hbox{\raise.25ex\hbox{$/$}\kern-.60em$\partial$}}
\newcommand{\beq}{\begin{equation}}
\newcommand{\eeq}{\end{equation}}
\newcommand{\ba}{\begin{eqnarray}}
\newcommand{\ea}{\end{eqnarray}}
\newcommand{\be}{\begin{equation}}
\newcommand{\ee}{\end{equation}}
\newcommand{\bea}{\begin{eqnarray}}
\newcommand{\eea}{\end{eqnarray}}

\usepackage[centertags]{amsmath}
\usepackage{amsfonts}

\begin{document}
\title{R-symmetry and Supersymmetry  Breaking at Finite Temperature}
\author{E.~F.~Moreno$^a$,
F.~A.~Schaposnik$^b$\thanks{Associated with CICBA}
\\
\\
{\normalsize $^a\!$\it Department of Physics,West Virginia University}\\
{\normalsize\it Morgantown, West Virginia 26506-6315, U.S.A.}
\\
{\normalsize $^b\!$\it Departamento de F\'\i sica, Universidad
Nacional de La Plata}\\ {\normalsize\it C.C. 67, 1900 La Plata,
Argentina}}
\maketitle
\begin{abstract}

We analyze the  spontaneous $U(1)_R$ symmetry breaking at finite
temperature for the simple O'Raifeartaigh-type model introduced in
\cite{Shih} in connection with spontaneous supersymmetry breaking.
We calculate the finite temperature effective potential (free
energy) to one loop order and study the thermal evolution of the
model. We find that the R-symmetry breaking occurs through a second
order phase transition. Its associated meta-stable supersymmetry
breaking vacuum is thermodynamically favored at high temperatures
and the model remains trapped in this state by a potential barrier,
as the temperature lowers all the way until $T=0$.

\end{abstract}
\newpage
\section{Introduction}
It became clear after the work of Nelson and Seiberg
\cite{NelsonSeiberg} that global $R$-symmetry plays a key role
in connection with supersymmetry breaking. In order to have
spontaneous supersymmetry breaking at the ground state of
generic models there must be a global $U(1)_R$ symmetry, but in
order to have non-zero gaugino masses it is necessary that this
symmetry be explicitly or spontaneously broken. The work of
Intriligator, Seiberg, and Shih (ISS) \cite{ISS} showed how
this tension between $R$-symmetry and supersymmetry can be
exploited to find generic models with an acceptably long lived
meta-stable supersymmetry breaking vacuum. Moreover, studying
the Seiberg dual of $ {\cal N} = 1$ super-QCD it has been shown
that, at high temperatures, the supersymmetry breaking vacua
are dynamically favored over the ``supersymmetry preserving''
ones\footnote{At finite temperature SUSY is always broken. With
the quotation marks we mean the phase which, for zero
temperature, corresponds to a supersymmetry preserving vacua.}
so that the Universe would naturally have been driven into them
\cite{ACJK}-\cite{J}, a possibility already discussed on
general grounds a long time ago in \cite{ELR}.

Different models with meta-stable symmetry breaking vacua and
structures rather different than those discussed by ISS have
been also investigated, as for example those based in gauge
mediation and extraordinary gauge mediation, which cover a
broad class of R-symmetric generic models with supersymmetry
breaking \cite{EOGM}-\cite{Essig}.

There is a very practical mechanism proposed in \cite{Shih}
leading to spontaneous $U(1)_R$ breaking. It applies  to
O'Raifeartaigh models with a continuous space of supersymmetry
breaking vacua and degenerate tree-level vacuum energy. It has
been shown in that work that, due to one loop corrections,
spontaneous $R$-symmetry breaking  occurs  \`a la
Coleman-Weinberg in a very simple O'Raifeartaigh type model and
for a wide range of parameters. More general models of this
kind have been discussed in \cite{KS} and their  thermal
history  has also been recently investigated \cite{Katz}.

It is the purpose of this work to study the question of
spontaneous $U(1)_R$ symmetry breaking at finite temperature
and the resulting supersymmetry breaking pattern  by analyzing
the thermal evolution of the O'Raifeartaigh-type model
introduced in \cite{Shih}. To this end we will compute the
finite temperature effective potential (i.e. the free energy
density) by shifting as usual the relevant background fields
and use the resulting quadratic terms (the mass terms) to
perform the one-loop calculation. Studying numerically the
corresponding one loop effective potential we will analyze the
nature of the different phase transitions,  showing how
parameters of the model can be chosen so as to cover the desire
range of critical temperatures at which R-symmetry breaking
takes place. As we shall see, our numerical results are
consistent with the general analysis presented in \cite{Katz}
where a broad class of models for gauge mediation were
considered. Indeed, in the classification of Extraordinary
Gauge Mediation Models (EOGM) of \cite{EOGM}, the model we
analyze belongs to the type I class (provided one promotes the
singlet messengers to fields transforming in the $5 \oplus 5$
representation of $SU(5)$). Our analysis will confirm the
thermal evolution scenario advanced in \cite{Katz} for type I
models, in particular concerning the existence of a metastable
vacuum at high temperatures with no $T=0$ analog

In the next section we introduce the model proposed in
\cite{Shih} and describe  its classical vacua, which includes a
moduli space and a runaway direction. We then  present  the
different terms that contribute to the one loop finite
temperature effective potential $V^1_\text{\it eff}$. In
section 3 we calculate  $V^1_\text{\it eff}$ along the
pseudo-modulus, which is at the origin of the dynamically
generated meta-stable vacuum, and analyze the R-symmetry
breaking phase transition. We then extend in section 4 the
calculation of $V^1_\text{\it eff}$ by considering a background
field that interpolates between the meta-stable vacuum  and the
runaway direction, and discuss in detail the resulting thermal
scenario. We finally summarize and discuss our results in
section 5.

\section{Set up of the model and the effective potential}
We consider the O'Raifeartaigh model for chiral superfields
considered in \cite{Shih},  with canonical K\"alher potential and
superpotential
\be W = \lambda X \phi_1\phi_2 + m_1 \phi_1\phi_3 +  \frac12
m_2\phi_2^2 + f X \label{super} \ee
This superpotential defines the underlying model which
communicates  supersymmetry breaking to the
minimal supersymmetric Standard Model. Chiral superfields
$\phi_i$ ($i=1,2,3$) with $R$ charges
\be R(\phi_1) = -1 \; , \;\;\;
R(\phi_2) = 1 \; , \;\;\; R(\phi_3) = 3 \; , \;\;\; \ee
represent the messengers of  supersymmetry breaking and the
spurion field $X$ generates the model's pseudo-moduli space and
has charge $R(X) = 2$. Parameters $\lambda$, $f$, $m_1$, and
$m_2$ will be taken, without loss of generality, as real
positive numbers.

The resulting scalar potential (we use the same notation for
superfields and their lowest components) takes the form
\be V^\text{\it tree}(X,\phi_i) = |\lambda \phi_1\phi_2 + f|^2
+ |\lambda X\phi_2 + m_1 \phi_3|^2 + |\lambda X \phi_1 +
m_2\phi_2|^2 + |m_1\phi_1|^2 \label{potencial} \ee
and its   extrema consist of:
\begin{itemize}
\item a moduli space
\be \phi_i^{(m)} = 0\; , \;\;\;\; X^{(m)} \;
\text{arbitrary}\label{cuatro} \ee
with \be V = f^2 > 0 \ee
\item a runaway direction
\begin{align}
&\phi_1^{(r)} = \left(\frac{f^2 m_2}{\lambda^2m_1\phi_3}\right)^{{\frac13}} \! , &&
\phi_2^{(r)} =- \left(\frac{fm_1\phi_3}{\lambda m_2} \right)^{{\frac13}} \! , &&
\phi_3^{(r)} \to \infty, \nonumber\\
&X^{(r)} = \left(\frac{m_1^2 m_2 \phi_3^2}{\lambda^2f} \right)^{{\frac13}}
\label{run}
\end{align}
with \be V \to 0\,. \ee
\end{itemize}
The moduli space does not correspond to  global minima  of the
potential but, as long as
\be |X| <
\frac{m_1}{\lambda} \frac{1 - y^2}{2y} \label{ocho} \ee
where
\be y = \frac{\lambda f }{m_1m_2} \label{y} \ee
it leads to local minima of the potential.  Since the $X$ field
is $R$-charged, such flat direction breaks the global $U(1)_R$
symmetry for any $X \ne 0$ in the range (\ref{ocho}). It is
clear now  that if quantum corrections  produce a minimum at
some point $\langle X \rangle \ne 0$ of this  flat direction,
which then corresponds to a pseudo-moduli, the associated
vacuum expectation value  will spontaneously break the
R-symmetry. This was shown at $T=0$ in \cite{Shih} by computing
the one-loop effective potential. We will now extend the
analysis to include thermal effects by computing the finite
temperature effective potential up to one loop, which takes the
form \cite{JD}
\be V^\text{\it eff}_1(X^{cl},\phi_{i}^{cl}) = V^\text{\it
tree}(X^{cl},\phi_{i}^{cl}) + V_1^0(X^{cl},\phi_{i}^{cl}) +
V_1^T(X^{cl},\phi_{i}^{cl})\,. \label{JD} \ee
The original fields are written in the form
\bea
X &=& X^{cl} +   x\nonumber\\
\phi_i &=& \phi_i^{cl} +   \varphi_i \label{deco} \eea
to proceed to compute the one-loop contribution by integrating
terms quadratic in the fluctuations $x, \varphi_i$. The zero
temperature piece $V_1^0 $ of the effective potential is given
by the usual supersymmetric generalization of the
Coleman-Weinberg formula
\be V_1^0 = \frac{1}{64\pi^2} {\cal S}\text{Tr} {\cal M}^4
\log\frac{{\cal M}^2}{\Lambda^2} \label{CW} \ee
where ${\cal S}\text{Tr}$ is the supertrace including a
negative sign for fermions,  ${\cal M}$ stands for the full
mass matrix resulting from the shift (\ref{deco}), ${\cal M}=
{\cal M}(X^{cla}, \phi_i^{cla})$, and $\Lambda$ is a mass
scale. Concerning the finite temperature contribution, one has
\cite{JD}
\be
 V_1^T = \frac{T^4}{2 \pi^2}
\sum_{i} \pm n_i \int_0^\infty ds \,s^2
\log\left(1\mp e^{-\sqrt{s^2+
  {\cal M}_{i}^2/T^2}}\,\right)
  \label{FT}
\ee
where the sum is over all degrees of freedom ($\{n_i\}$ denotes
the number of degrees of freedom, $n=2$ for complex scalars and
Weyl fermion  and the upper (lower) sign is for bosons
(fermions)). Finally, ${\cal M}_i$ denotes the eigenvalues of
the ${\cal M}$-matrix.

In order to make contact between the parameters of the model
with scalar potential (\ref{potencial}) and those of the
Minimal Supersymmetric Standard Model (MSSM) one has to
consider  masses of the observable fields. It should be
mentioned that a superpotential of the type (\ref{super})
should be in principle supplemented  with a minimal gauge
mediation (MGM) messenger $\phi_4$  which, coupled to the
spurion field $X$ through a term  of the form $X \phi_4^2$,
will effectively give a mass to the otherwise massless gaugino
\cite{EOGM}. Note that the introduction of this additional
messenger would promote our model to a type III one, for which,
instead of a condition of the form (\ref{ocho}) stability
requires an upper and a lower bound for $X$, $X_\text{max}
> |X| > X_\text{min}$, as noted in \cite{EOGM} for $T=0$ and
discussed in \cite{Katz} for finite $T$. In the case of the
model we consider one should adjust the parameters so that such
bounds hold at all temperatures and as the temperature grows
$X_\text{min}(T)$ approaches the origin faster than the
pseudomodulus minimum. We leave for a future work a detailed
analysis of this issue and proceed to determine the orders of
magnitude of the different superpotential parameters by
analyzing sfermion masses.

Sfermion masses $m_{\tilde f}^2$ can be extracted from the
matter wave function renormalization through the formula
\cite{GR},
\be
m_{\tilde f}^2 \sim \frac{\alpha^2}{(4\pi)^2}
\left(\frac{f}{m}\right)^2 {\tilde N}
\label{mm}
\ee
where $\alpha$ is the running coupling constant of the
underlying gauge theory (evaluated at the messenger scale,
$\alpha/4\pi \sim 10^{-2}$) and
\be \tilde N = \lambda^2 \frac{\partial^2}{\partial x \partial
{x^*}} \sum_{i=1}^3 \log^2 |{\cal M}_{Fi}|^2\,. \ee
Here ${\cal M}_{Fi}$ are the eigenvalues of the fermion mass
matrix  resulting from superpotential (\ref{super}) and for
simplicity we have  set $m_1 = m_2 = m$ and defined $x =
\lambda X / m$. Given configuration (\ref{cuatro}),  ${\cal
M}_F$ can be written in the form
\be
{\cal M}^2_F= m^2 \left(\begin{matrix}x x^* + 1 & x & 0 \\
x^* & x x^* + 1 & x \\ 0 & x^* & 1\end{matrix}\right)
\ee
Formula (\ref{mm}) is valid in the regime $f \ll m^2$ for which
supersymmetry is broken only in the effective field theory
bellow the messenger scale by soft terms.

Now one can check that
\be
\tilde N (x\to 0) = \lambda^2 \; , \;\;\;\;
\tilde N (x \to \infty) = 0
\ee
Moreover, had we added the MGM messenger, the $\tilde N$
behavior at infinity would have raised to $\lambda^2$ so that
we can take $\tilde N \sim \lambda^2$ in the whole range. In
fact, if one scales $X \to X/\lambda$ and $f \to \lambda f$ the
coupling $\lambda$ completely disappears from superpotential
(\ref{super}) so that we can just set  $\tilde N \sim 1$ in
(\ref{mm}).

Since one expects that the sfermion mass should be in the TeV
scale, one infers from (\ref{mm}) that $f/m \sim 100$ TeV, this
in turn implying that $m \gg 100$ TeV. The estimate would
remain nearly unchanged if instead of the assumption $f \ll
m^2$ we consider the case $f\sim m^2$.  We   conclude that for
the analysis of the  thermal evolution of the system,  high
temperatures will correspond to $T\gg 100$ TeV.

\section{The fate of the meta-stable vacuum}
We start by considering the effective potential  for
configuration (\ref{cuatro}), that is, we take $\phi_i^{cl} =
\phi_i^{(m)} = 0$ and $X^{cl} = X^{(m)} = X$ in formula
(\ref{JD}). In this case the boson  mass matrix takes the form
(we omit the superscript $m$)
\be
{\cal M}^2_B = \left(
\begin{array}{llllll}
m_1^2+\lambda ^2 X^2 & m_2 \lambda  X
& 0 & 0 & f \lambda & 0 \\
m_2 \lambda  X   & m_2^2+\lambda ^2 X^2 & m_1 X
\lambda  & f \lambda  & 0 & 0 \\
0 & m_1 \lambda  X   & m_1^2 & 0 & 0 & 0 \\
0 & f \lambda  & 0 & m_1^2+\lambda ^2 X^2 & m_2 \lambda  X
& 0 \\
f \lambda  & 0 & 0 & m_2 \lambda  X   & m_2^2+\lambda ^2
X^2  & m_1 \lambda  X   \\
0 & 0 & 0 & 0 & m_1 \lambda  X   & m_1^2
\end{array}
\right)
\ee
while the fermion mass matrix reads
\be
{\cal M}^2_F =
\left(
\begin{array}{llllll}
m_1^2+\lambda^2 X^2 & m_2 \lambda X & 0 & 0 & 0 & 0 \\
m_2 \lambda X & m_2^2+\lambda^2 X^2 & m_1 x  \lambda  & 0 & 0 & 0 \\
0 & m_1 \lambda X & m_1^2 & 0 & 0 & 0 \\
0 & 0 & 0 & m_1^2+\lambda^2 X^2 & m_2 \lambda X & 0 \\
0 & 0 & 0 & m_2 \lambda X & m_2^2+\lambda^2 X^2 & m_1 \lambda X \\
0 & 0 & 0 & 0 & m_1 \lambda X & m_1^2
\end{array}
\right)
\label{masaf}
\ee
Using this result, one can compute the zero-temperature
one-loop contribution (\ref{CW}), as originally calculated in
\cite{Shih},
\be V_1^0 = \frac{1}{64\pi^2} {\rm Tr} \left( {\cal M}_B^4
\log\frac{{\cal M}_B^2}{\Lambda^2} - {\cal M}_F^4
\log\frac{{\cal M}_F^2}{\Lambda^2} \right) \label{veinte} \ee
as well as the finite temperature one, eq.(\ref{FT}), which can
be  rewritten in the form \be {V_1}^T = \frac{T^4}{2 \pi^2}
\sum_{i=1}^6 \int_0^\infty ds \,s^2\left( \log(1-e^{-\sqrt{s^2+
{\cal M}_{Bi}^2/T^2}}) - \log(1+e^{-\sqrt{s^2+   {\cal
M}_{Fi}^2/T^2}}) \right) \label{veintiuno} \ee

One can scale $X \to m_1 X/\lambda$ and masses so that the
effective potential only depends on the rescaled $X$ and on two
parameters: $y$, defined in eq. \eqref{y}, and $r$, given by
\be r = \frac{m_2}{m_1} \ee so that $V^\text{\it eff}_1 =
V^\text{\it eff}_1(X;r,y)$ with $m_1$ giving the mass scale.

Eigenvalues ${\cal M}_{Bi}$ and ${\cal M}_{Fi}$ (with
$i=1,\ldots,6$) of mass matrices ${\cal M}_B$ and ${\cal M}_F$
have to be computed numerically.  Of course, at $T =0$ one
reproduces the results in \cite{Shih} thus finding that, for a
wide range of parameters, there is a meta-stable  vacuum where
$U(1)_R$ is spontaneously broken. Concerning the thermal
evolution we show in figure 1 the plot of $V^\text{\it eff}_1$
as a function of $X$ for different temperatures. In figure 2 we
represent the change with temperature of the region (shown in
white) in the $r,y$ plane where there is a $U(1)_R$ symmetry
breaking local minimum of the potential satisfying
(\ref{ocho}).

\vspace{1 cm}

\centerline{\epsfxsize=4. in \epsffile{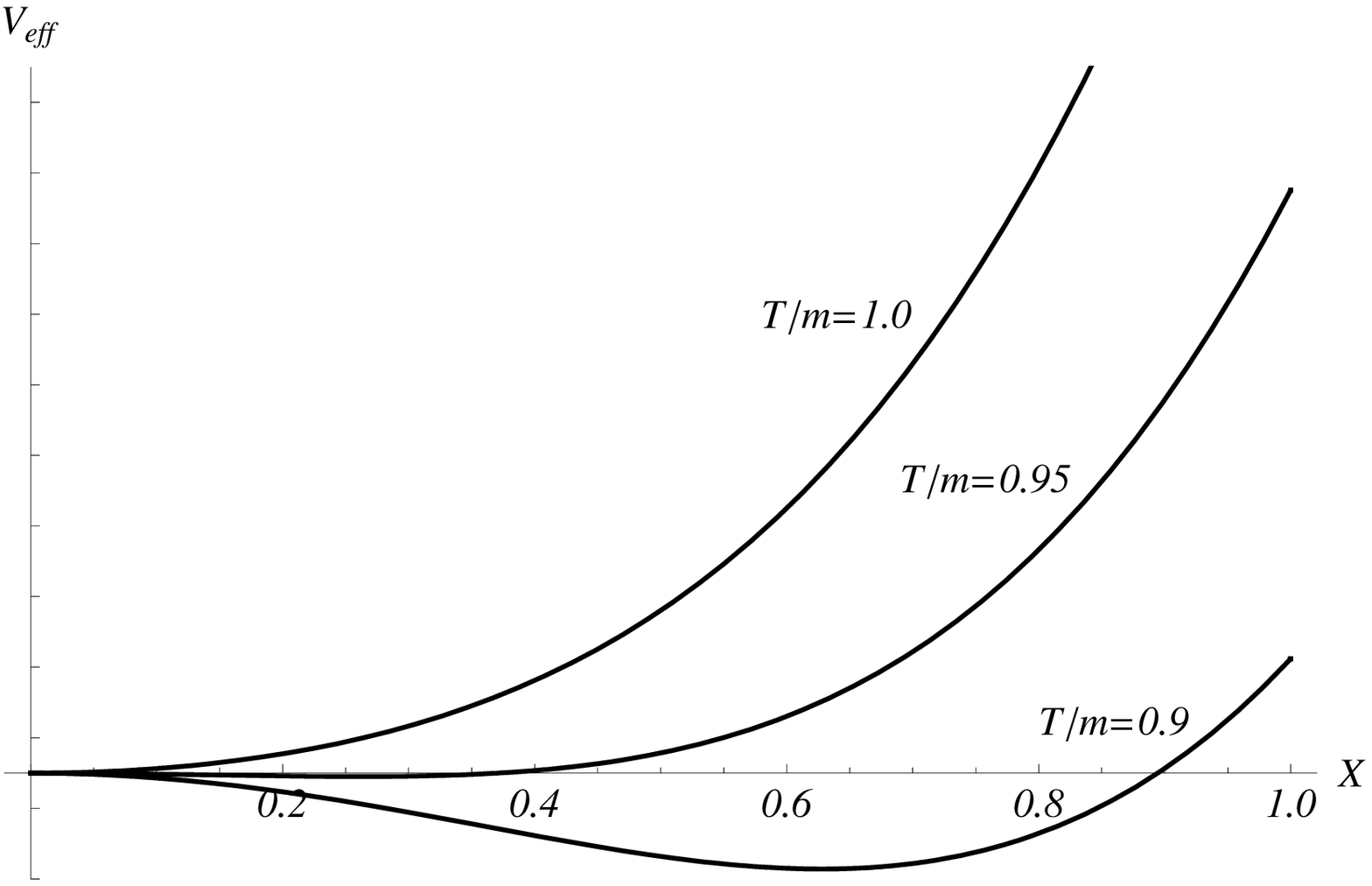}}

\noindent Figure 1: The effective potential as a function of
$|X|$ showing the second order phase transition (we have taken
$r=4$ and $y= 0.2$). The curve in the middle corresponds to the
critical temperature which for the chosen parameters takes the
value $T_R/m = 0.95$.

~

\vspace{1.4 cm}

\centerline{\epsfxsize=6. in \epsffile{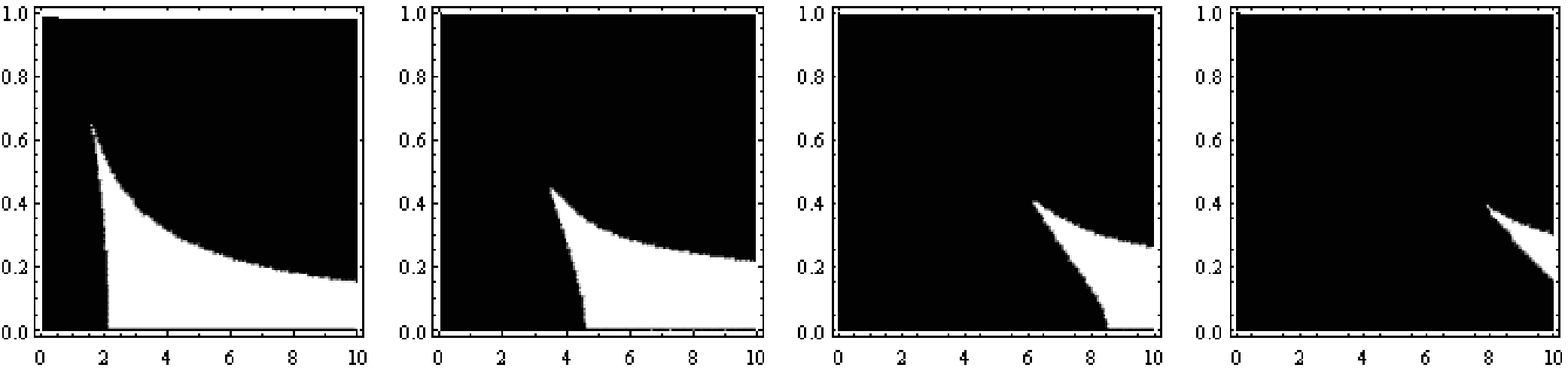}}

\noindent Figure 2: Plot of $y = \lambda f/(m_1m_2)$ as a
function of  $r=m_2/m_1$ for $T=0, 1, 1.5$, and $1.8$ (from
left to right). The white region corresponds to a local
(R-symmetry breaking) minimum (with no tachyons).

 ~

Using different pairs of values $(r,y)$ in the range where $R$
symmetry breaking occurs (white region in Figure 2) we have
then found  a second order phase transition at a certain
critical temperature $T_{R}$,  so that for  $T<T_R$ there is a
minimum away from the origin, i.e. at $ X = \langle X \rangle
\ne 0$.

Interestingly enough, changing parameters one can make the
critical temperature vary in a wide range. For example, for the
choice of parameters corresponding to Figure 1, ($y = 0.2,
r=4$) the critical temperature is $T_{R}/m = 0.95$ while for
$y=0.2, \, r=2.07$ it becomes $T_{R}/m \sim 10^{-3}$. In fact,
by choosing parameters $(r,y)$ closer and closer to the left
frontier of the white region in Figure 2 one can lower the
critical temperature as much as wanted. Taking into account the
condition $ m \gg 100$ TeV previously found from the
requirement that $m_{sf} \sim 1$ TeV, we see that the critical
temperature at which R-symmetry is broken can be adjusted in a
wide range going for the two choices we have used as example,
from $T_{R} \gg 100$ TeV to $T_{R} \sim 1$ TeV. It should be
noted that as the value of the critical temperature  lowers the
R-symmetry breaking VEV $\langle X \rangle$ gets closer to the
origin.

\section{The fate of the runaway direction}
We will now study the behavior of the runaway direction as the
temperature changes. To this end, we will follow  an approach
similar to that used in \cite{ACJK} in the case of the ISS and
consider a path $(X^\text{\it int}, \phi_i^\text{\it int} )$
interpolating between the meta-stable supersymmetry vacua and
the supersymmetric runaway direction. A convenient choice of
path is
\[
X^\text{\it int} = \left(\frac{m_1^2 m_2 \phi_3^2}{\lambda^2f} \right)^{{\frac13}}
+ \left( 1 - h(\phi_3)\right) \langle X \rangle\; ,
\]
\be \phi_1^\text{\it int} = h(\phi_3)  \left(\frac{f^2
m_2}{\lambda^2m_1\phi_3}\right)^{{\frac13}} \; , \;\;\;\;
\phi_2^\text{\it int} = - \left(\frac{fm_1\phi_3}{\lambda m_2}
\right)^{{\frac13}}  \; , \;\;\;\; \phi_3^\text{\it int} =
\phi_3, \;\;\;\; \label{path} \ee
The function $h(\phi_3)$ should be chosen so as to conciliate
the behavior of $X$ and $\phi_1$ at the two-endpoints. An
appropriate election is
\be
h(y) =\frac{2}{\pi} \arctan cy
\ee
where $c$ is a parameter to be chosen so that the path, which
goes from the zero temperature meta-stable local minima
($\phi_3=0$) at $X = \langle X \rangle $ to the runaway value
($\phi_3 \to \infty$) does not have modes with negative square
masses.

We present in an Appendix the explicit form of   boson and
fermion masses for the path (\ref{path}). From their explicit
form  one can  numerically study the effective potential as a
function of $\phi_3$ and the temperature, $V^\text{\it eff}_1 =
V^\text{\it eff}_1(\phi_3,T)$, and determine the resulting
minima landscape. First, one has to numerically compute  the
mass eigenvalues and then  evaluate the zero temperature
one-loop contribution to the effective potential
(eq.(\ref{veinte}))  as well as the finite temperature one,
$V_1^T$, given by eq.(\ref{veintiuno}).

One should note that at very high temperatures  $V_1^T$, as
given by formula (\ref{FT}),  becomes
\be V_1^T \sim
-\frac{\pi^2}{8} T^4 \;\;\;\; \;\;\;\; {\rm for~} T \to \infty
\ee
Note that the negative sign in the effective potential is
harmless since at finite temperature $V^\text{\it eff}_1$
should be identified  with the free energy as a function of the
order parameter while the total energy is given by
\be E = V^\text{\it eff}_1 - T \frac{\partial V^\text{\it
eff}_1}{\partial T} \ee
which is indeed positive for all temperatures. We show in
figure 3 the free energy $V^\text{\it eff}_1$ (left) and the
total energy $E$ (right) at very high temperatures. The figure
clearly shows  that although the energy is lower in what will
become at zero temperature the runaway direction, the entropy
contribution favors the non supersymmetric free energy minimum
near the origin

~

\centerline{\epsfxsize=5.9 in \epsffile{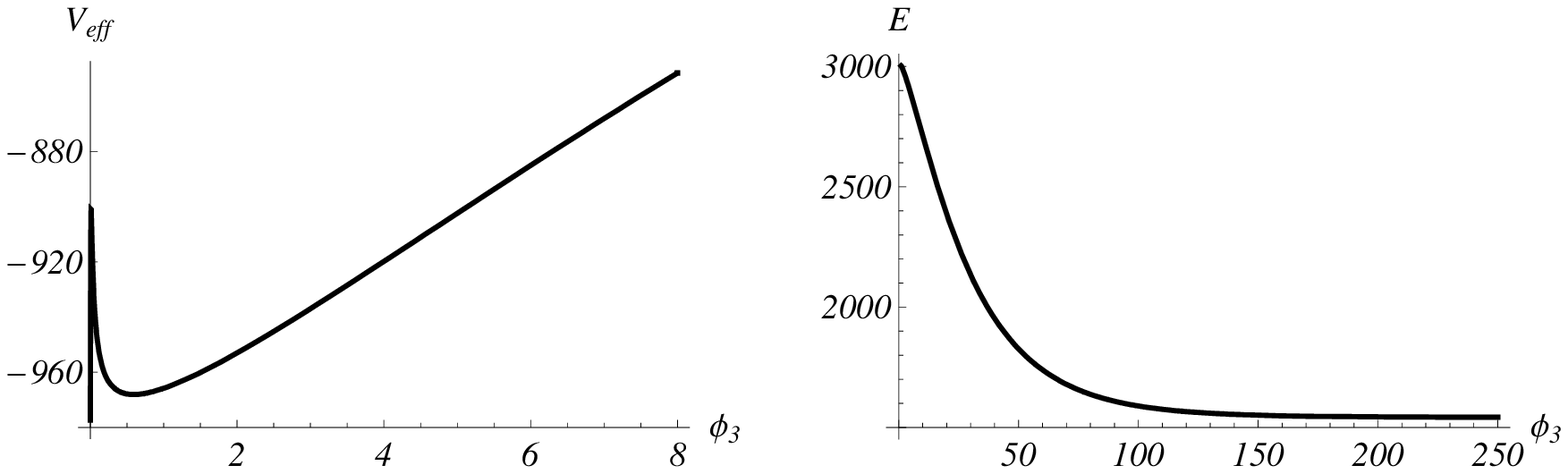}}

\centerline{Figure 3: Free energy vs. total energy for
$T/m=5$.}

\vspace{0.7 cm}

From the numerical analysis of the complete effective potential
$V^\text{\it eff}_1(\phi_3,T)$ one infers the following
scenario for the thermal evolution of the effective potential:

\begin{itemize}
\item For $T/m \gg 1$ the potential has an absolute minimum
    at the origin in field space and it grows without bound
    for large values of $\phi_3$. The zero-temperature
    meta-stable vacuum in the pseudomoduli direction has
    not yet started to develop and one finds, in addition,
    a local minimum at at a finite value $\phi_3^*$ (i.e.
    $V^{\text{\it eff}\,*}_1(\phi_3^*, T^*) > V^{\text{\it
    eff}\, *}_1(0,T^*)$)
\item  As the temperature lowers, the slope of the
    potential at infinity decreases until it becomes
    negative. The change of sign takes place at a
    temperature $T_h$ at which the absolute minimum of the
    potential is still at the origin.
 \item  At a lower temperature $T_b$ the local minimum
     $V^{\text{\it eff}\, *}_1$ disappears.
 \item At a lower temperature  $T_{ra}$, $V^\text{\it
     eff}_1({\phi_3 \to \infty}, T_{ra})  = V^\text{\it
     eff}_1(0,T_{ra})$  so that
   the runaway minimum appears and a first order phase
   transition starts.
 \item  As already discussed, at a lower temperature
     $T_{R}$ the $R$-symmetry breaking meta-stable vacuum
     arises.
\end{itemize}
As an example, for the parameter choice $r=4, y= 0.2$ already
used to discuss the meta-stable vacuum evolution, the
temperatures defined above take the values
\be T_h/m = 2.96 \;
, \;\;\;\; T_b/m = 1.29 \; , \;\;\;\; T_{ra}/m = 1.14 \; ,
\;\;\;\; T_{R}/m =  0.95
\ee

We have already described how changing parameters $(r,y)$
towards the left border of the $R$-symmetry braking region
(white region in Fig.2) lowers the critical temperature at
which the transition to the meta-stable vacuum takes place. All
other temperatures lower but their change is not so marked. As
an example, for $(r=2.7 , y=0.2)$ one has
\be
T_h/m = 1.5 \; , \;\;\;\;
T_b/m = 0.99 \; , \;\;\;\;
T_{ra}/m = 0.81 \; , \;\;\;\;
T_{R}/m = 1\times 10^{-3}
\ee
Figure 4 shows a qualitative representation of the above
scenario.

\centerline{\epsfxsize=3.9 in \epsffile{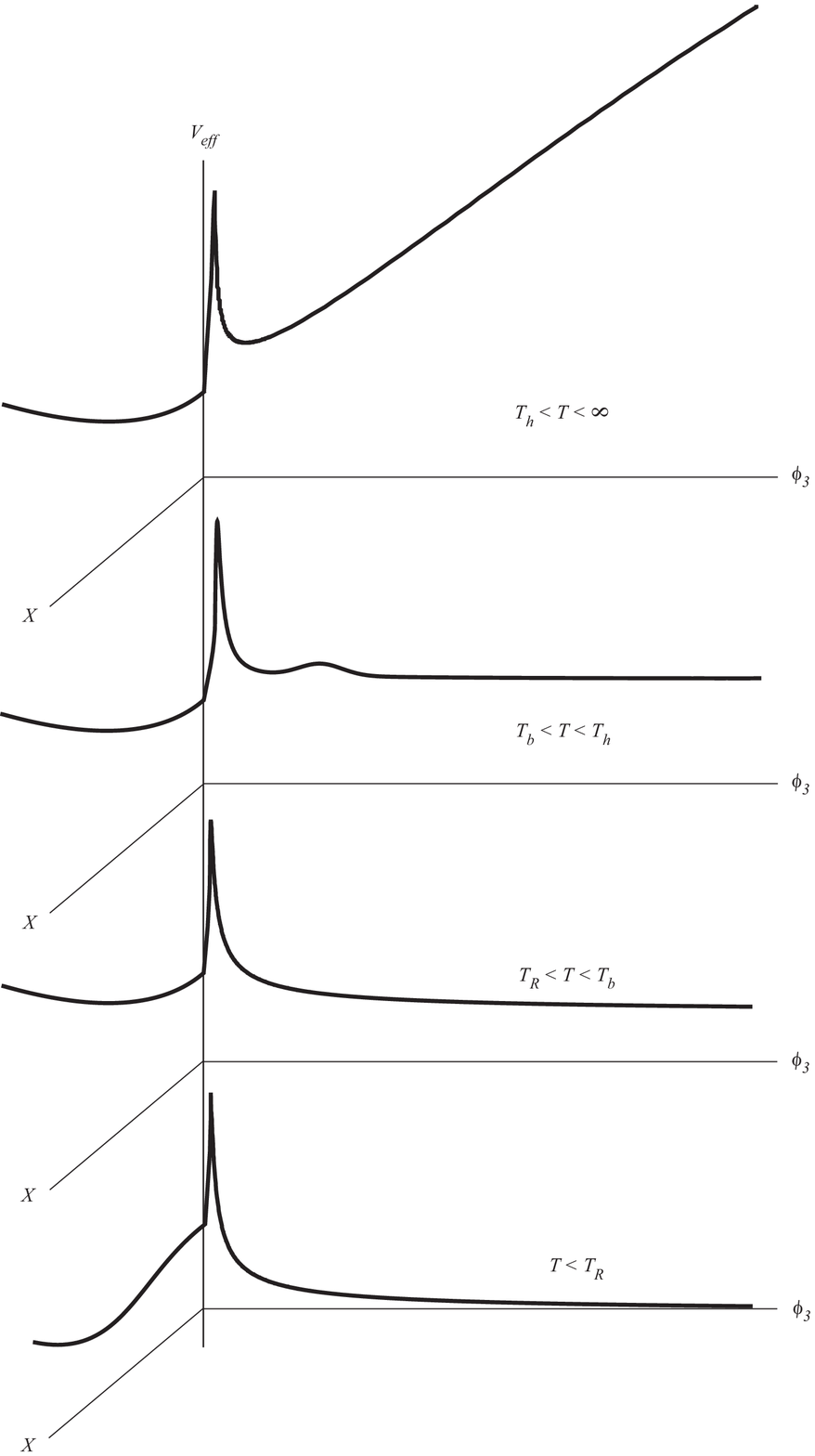}}

\centerline{Figure 4: Evolution of the effective potential with
temperature}

\vspace{ 0.5 cm}

In order to exclude the  possibility that the system escapes
towards the runaway direction instead of decaying into the
meta-stable vacuum  let us note that for $T > T_b$ the
effective potential has an absolute minimum at the origin. Only
for temperatures  $T \leq T_b$ the runaway direction
corresponds to an (asymptotic) global minimum of the effective
potential. Since such temperatures are sufficiently low as to
neglect thermal corrections, one can see \cite{Shih} that the
barrier preventing the system to roll-down along the runaway
direction has a width of order $y^{-1}$ while its height is of
order $y^0$. Hence, by taking $y$ sufficiently small the system
will remain in the vacuum at the origin while $T_b<T<T_R$ and
then smoothly evolve towards the meta-stable vacuum for
$T<T_R$.

\section{Discussion}
We have analyzed the thermal evolution of the simplest
O'Raifeartaigh-type model in which spontaneous R-symmetry
breaking occurs dynamically, leading to a runaway behavior at
large fields and a meta-stable vacuum which, at zero
temperature, spontaneously breaks  supersymmetry. Studying the
effective potential at finite temperature we have shown that
the $U(1)_R$ breaking arises through a second order phase
transition. Remarkably, the critical temperature at which the
R-symmetry breaking phase starts can be lowered by an
appropriate choice of parameters and this also implies that the
VEV of the spurion field $X$ also decreases.

We also analyzed the thermal evolution of the runaway direction
finding, as expected, that high temperature contributions rise
the asymptotic directions of the effective potential.
Remarkably, we found that at high temperatures there is an
extra local minimum of the effective potential, though
energetically unfavored with respect to the meta-stable vacuum.
At some temperature ($T_b$) this local minimum disappears.

The whole thermal evolution sequence is as follows: At high
temperatures the model is driven to the meta-stable
SUSY-breaking vacuum. As the temperature decreases, the SUSY
runaway direction becomes energetically favored but the
transition between phases is long lived, so the system remains
in the meta-stable vacuum. There is also an extra local minimum
but with higher effective potential than the meta-stable
vacuum. As the temperature decreases this extra minimum fades
away. Finally, at an even lower temperature ($T_R$), the R
symmetry is broken and a second-order phase transition occurs.
This sequence, with the exception of the existence and eventual
disappearance of the extra local minimum, is similar to the one
described in \cite{ACJK}-\cite{K} for the magnetic dual of
SuperQCD. As stated in the introduction, the model studied here
can be extended to the form of a type I model in the
classification of ref.\cite{EOGM}. The general properties of
the thermal evolution of these models was discussed in
\cite{Katz} and our numerical analysis of the vacuum structure
at different temperatures is consistent with them. In
particular our results confirm the existence of an extra vacuum
at high T in addition to the one at the origin, with no analog
at $T=0$. This extra vacuum disappears as the temperature
lowers below $T_b$.

~

An implicit assumption necessary to apply our results in a
cosmological context is that the reheat temperature $T_{\rm
reheat}$ is larger enough (with respect to the supersymmetry
breaking scale) as to guarantee that the supersymmetry breaking
history develops quasi-statically, in a situation of thermal
equilibrium. This justifies to look for the minima  of  the
free energy  not taking into account possible interaction
between fields and the heat bath. Ignoring the possibility of
non-equilibrium situations  our results suggest that although
the runaway direction starts to develop before the R-symmetry
breaking meta-stable minimum appears, the system will not
roll-down from the minimum at the origin because of the
existence of a very high barrier so that when the $R$- symmetry
breaking meta-stable vacuum is available, it will evolve to it
and remain there for a sufficiently large time as to ensure
that the Universe is  still  trapped there.

We would like to end this work by pointing out two directions
in which we hope to continue our investigation on R-symmetry
breaking and supersymmetry breaking at finite temperature. One
concerns the analysis of models with explicit R-symmetry
breaking which, under certain conditions, have supersymmetric
vacua, runaway directions and meta-stable vacua \cite{MS}. As
discussed in \cite{Abel}, the way in which R-symmetry is broken
(spontaneously or explicitly) leaves a clear imprint on the
phenomenology of the MSSM and it is then worthwhile to study
broad classes of such models so as to compare the resulting
thermal patterns. The other direction is related to the
analysis in \cite{Shih2} on how pseudomoduli arising in
generalized O'Raifeartaigh models from additional global
symmetries can be candidates to dark matter (see also
\cite{KerenZur}). In this context it would be of interest to
investigate the thermal evolution of such models along the
lines developed here. We hope to analyze these issues in a
future work.

\vspace{1 cm}

\noindent\underline{Acknowledgments}  We would like to thank
Diego Marqu\'es for his criticism and helpful comments. This
work was partially supported by  PIP6160-CONICET,  BID
1728OC/AR PICT20204-ANPCYT grants and by CIC and UNLP,
Argentina.

\newpage
\section{Appendix}
We write the boson and fermion mass matrices corresponding to the path
(\ref{path}) in the form
\be
{\cal M}_F^2 =
\left(
\begin{array}{ll}
A & 0 \\
0 & A
\end{array}
\right)
\; , \;\;\;\;\;
{\cal M}_B^2 = \left(\begin{array}{ll}
A & B \\
B & A
\end{array}
\right)
\ee
where $A$ and $B$ are symmetric $4\times4$ matrices with nonzero elements
\bea
A_{11} &=& \frac{r^2 y^{4/3} h(\phi_3 )^2 m_1 ^{8/3}}{\phi_3 ^{2/3}
 \lambda ^{2/3}}+y^{2/3} \phi_3 ^{2/3} \lambda ^{2/3} m_1 ^{4/3} \nonumber\\
 A_{12} &=& A_{21} = \frac{m_1 ^{4/3} r y^{2/3} \lambda ^{2/3} h(\phi_3 ) \left(-h(\phi_3 )
   x_0+x_0+\frac{\sqrt[3]{m_1 } \phi_3 ^{2/3}}{\sqrt[3]{y}
   \sqrt[3]{\lambda }}\right)}{\sqrt[3]{\phi_3 }}
 \nonumber\\
 A_{13} &=& A_{31} = -m_1 ^{2/3} x_0 \sqrt[3]{y} \sqrt[3]{\phi_3 } \lambda ^{4/3}-m_1  \phi_3  \lambda
   +\frac{\left(r^2 y^{2/3} m_1 ^{7/3}+x_0 \sqrt[3]{y} \phi_3 ^{2/3}
    \lambda ^{5/3} m_1 ^{2/3}\right) h(\phi_3 )}{\sqrt[3]{\phi_3 } \sqrt[3]{\lambda }}
    \nonumber\\
    A_{14} &=& A_{41} = -m_1 ^{5/3} \sqrt[3]{y} \sqrt[3]{\phi_3 }
   \sqrt[3]{\lambda } \nonumber\\
 A_{22} &=&  m_1 ^2+y^{2/3}
   \phi_3 ^{2/3} \lambda ^{2/3} m_1 ^{4/3}+\left(x_0 \lambda -x_0 h(\phi_3 ) \lambda +
   \frac{\sqrt[3]{m_1 } \phi_3 ^{2/3} \lambda ^{2/3}}{\sqrt[3]{y}}\right)^2
   \nonumber\\
   A_{23} &=&    A_{32} =
      m_1  r \left(x_0 \lambda
   +\frac{\sqrt[3]{m_1 } \phi_3 ^{2/3} \lambda ^{2/3}}{\sqrt[3]{y}}-
   (m_1  y+x_0 \lambda ) h(\phi_3 )\right) \nonumber\\
A_{33} &=&    \frac{r^2 y^{4/3}
   h(\phi_3 )^2 m_1 ^{8/3}}{\phi_3 ^{2/3} \lambda ^{2/3}}+r^2 m_1 ^2+
   \left(x_0 \lambda -x_0 h(\phi_3 ) \lambda +
   \frac{\sqrt[3]{m_1 } \phi_3 ^{2/3} \lambda ^{2/3}}{\sqrt[3]{y}}\right)^2
   \nonumber\\
   A_{34} &=& A_{43} =
    \frac{\phi_3 ^{2/3} \lambda ^{2/3} m_1 ^{4/3}}{\sqrt[3]{y}}+
   x_0 \lambda  m_1 -x_0 \lambda  h(\phi_3 ) m_1 \nonumber\\
   A_{44} &=& m_1^2 \nonumber \\
   B_{12}&=& B_{21} =  \frac{m_1 ^{4/3} r \sqrt[3]{y} \sqrt[3]{\lambda } (h(\phi_3 )-1)
  \left(\sqrt[3]{m_1 } \phi_3 ^{2/3}-
  x_0 \sqrt[3]{y} \sqrt[3]{\lambda } h(\phi_3 )\right)}{\sqrt[3]{\phi_3 }}\nonumber\\
  B_{13} &=& B_{31} = m_1 ^{2/3}
   x_0 \sqrt[3]{y} \sqrt[3]{\phi_3 } \lambda ^{4/3} (h(\phi_3 )-1) \nonumber\\
  \nonumber\\
   B_{23} &=& B_{32} = -m_1 ^2 r y
   (h(\phi_3 )-1)
 \eea



\end{document}